\documentclass[floatfix,twocolumn,aps,prl,draft,showpacs]{revtex4}

\usepackage{amssymb}
\usepackage{graphicx}

\newcommand{\bu}{{\bf  u}}
\newcommand{\bx}{{\bf  x}}

\newcommand{\de}{{\mathrm  d}}

\newcommand{\BE}{\begin{equation}}
\newcommand{\EE}{\end{equation}}
\newcommand{\BA}{\begin{array}}
\newcommand{\EA}{\end{array}}
\newcommand{\beqn}{\begin{eqnarray}}
\newcommand{\eeqn}{\end{eqnarray}}
\newcommand{\nablab}{\mbox{\boldmath $\nabla$}}
\newcommand{\thetaxt}{\theta(\bx,t)}
\newcommand{\txt}{\theta(x,t)}

\begin{document}
\title{Combustion dynamics in steady compressible flows}
\author{S. Berti$^1$, D. Vergni$^2$ and A. Vulpiani$^3$}

\address{
$^1$ Laboratoire de Spectrom\'etrie Physique, Universit\'e Joseph Fourier 
Grenoble I and CNRS, BP~87, 38402 Saint Martin d'H\`eres, France.\\
$^2$ Istituto Applicazioni del Calcolo (IAC) - CNR, Viale del Policlinico, 137, 
I-00161 Roma, Italy.\\
$^3$ Dipartimento di Fisica, CNISM, and INFN Universit\`a di Roma ``La Sapienza'',
P.zle Aldo Moro 2, I-00185 Rome, Italy.
}

\begin{abstract}
We study the evolution of a reactive field 
advected by a one-dimensional compressible velocity field
and subject to an ignition-type nonlinearity.
In the limit of small molecular diffusivity the problem can be
described by a spatially discretized system, and this allows
for an efficient numerical simulation. 
If the initial field profile is supported in a region of size
$\ell < \ell_c$ one has quenching, {\it i.e.}, flame extinction,
where $\ell_c$ is a characteristic length-scale depending 
on the system parameters (reacting time,
molecular diffusivity and velocity field).
We derive an expression for $\ell_c$ in terms of these
parameters and relate our results to those 
obtained by other authors for different flow settings.
\end{abstract}

\pacs{47.70.Fw, 05.60.-k, 82.20.-w}
\maketitle

Front propagation in reaction-transport systems is a widely studied
topic in both scientific and applicative fields such as the dynamics
of biological populations, chemical reactions in fluids and flame
propagation in gases~\cite{Williams,Abraham,Epstein,rmv88}.

From the mathematical point of view, these phenomena can be modeled in
terms of partial differential equations describing the evolution of
both the concentrations of the reacting species, and the velocity
field~\cite{Peters,xin2000}.  Though in principle these equations are
coupled, a simplification comes from the assumption of no
back-reaction of the reactants concentration on the velocity
field. In this passive limit one can use an
advection-reaction-diffusion equation. The most compact model
considers the evolution of a single scalar field $\thetaxt$
representing the fractional concentration of products, or a normalized
temperature in the case of combustion processes, taking values in the
interval $[0,1]$.

The interest, and the difficulty, in the treatment of this subject is
due to the effect of advection on the reaction process: theoretical
studies~\cite{Constantin1,Pomeau,Yakhot}, numerical
simulations~\cite{Constantin2,acvv01,kn04} and laboratory
experiments~\cite{Bradley,srby92} show that the propagation speed of
the front is significantly altered by the presence of the fluid flow.
When an infinite reservoir of inert material is present, advection
enhances the speed of travelling waves. On the other hand, if the
initial condition is localized in a region of finite size, for a
certain class of reaction dynamics, the combined action of diffusion and
advection might reduce and eventually suppress front propagation. It
is then interesting to study how the critical size of the initial
support, below which the reactive process quenches, depends on
the characteristics of both the velocity field and the reaction 
dynamics~\cite{Constantin1,Constantin2}.

In this letter we study the quenching phenomenon, or flame extinction
in combustion terminology, in a one-dimensional compressible velocity
field in the limit of small molecular diffusivity. The reactive
dynamics is modeled by means of an ignition-like nonlinearity, that is
a reaction term with a threshold value $\theta_c$, such that if
$\theta<\theta_c$ no reaction takes place.  We derive a relation
between the critical size of the initial condition width and the
relevant parameters of the problem, namely the reaction time, the
reaction threshold value and the combined effect of diffusivity and
flow intensity. In the end we will compare our results with those
obtained by other authors in different contexts, {\it i.e.}, reactive
field advected by bidimensional incompressible velocity
fields~\cite{Constantin1}.

\section{Model}
Consider the usual advection-reaction-diffusion problem
\begin{equation}
\partial_t \theta + \nablab \cdot \left (\bu \, \theta \right )=
D_0\nabla^2\theta+{1 \over \tau}f(\theta),
\label{eq:ard}
\end{equation}
where $\bu(\bx,t)$ is a given compressible velocity field, $D_0$ is
the molecular diffusivity and $f(\cdot)$ the reactive term with its
characteristic time $\tau$. For the sake of simplicity we adopt a
one-dimensional stationary model with velocity field:
\begin{equation}
u(x)=U_0 \sin \left({\pi x \over L}\right)\,.
\label{eq:vcompress}
\end{equation}

Let us first discuss the system dynamics in absence of reaction.
The Lagrangian equation
\begin{equation}
{\de x \over \de t} = u(x)
\label{eq:lagrangian}
\end{equation}
has the following stable fixed points (for $U_0 > 0$) 
$x=\pm L, \pm 3L, \ldots, \pm(2n-1)L, \ldots$ 
while $x=0, \pm 2L, \ldots, \pm 2n L, \ldots$ are unstable.  
In absence of reaction the field $\theta$ will concentrate around 
the stable fixed points $x_n=(2n-1)L$ with $n = 0, \pm 1, \pm 2,
\ldots$ and, essentially, one has a random walk among the points $x_n$.
The characteristic time of jumping is determined by $U_0$ and $D_0$.

In a suitable range of values of $U_0$ and $D_0$ the field
$\theta(x,t)$ is well peaked around $x_n$, so 
we can introduce the variable $\theta_n$:
\begin{equation}
\theta_n(t) = \int_{x_n-L}^{x_n+L} \txt \de x 
            = \int_{x_n-\delta}^{x_n+\delta} \txt \de x \,,
\label{eq:thetan}
\end{equation}
where $\delta \ll L$. It is not difficult to write down
the evolution equation for $\theta_n(t)$:
\begin{equation}
\theta_n(t + \Delta t) = \sum_j P_{j\to n}^{(\Delta t)} \theta_j(t)
\label{eq:evt}
\end{equation}
where 
\begin{equation}
P_{n\to n}^{(\Delta t)} = 1 - 2W\Delta t \qquad 
P_{n\to n-1}^{(\Delta t)} = P_{n\to n+1}^{(\Delta t)} = W\Delta t ,
\label{eq:defP}
\end{equation}
and $W$ is a function of $U_0$ and $D_0$, {\it i.e.}, the escape rate of
a Brownian particle from a potential well. 
For small $D_0$ it is possible to show that $\ln W \sim -{U_0 \over
D_0}$, which is the celebrated Kramers formula~\cite{Gardiner}.  
For generic $D_0$ and periodic velocity field $u(x)$ it is not 
difficult to have good numerical estimate of $W$.

In equation~(\ref{eq:evt}) both time and space are discrete.
However, while the time discretization is merely due to numerical
reasons, the discretization of space is a consequence
of compression, and in the limit of small $D_0$ (and not small $U_0$)
equation~(\ref{eq:evt}) 
is a very good approximation. 
It is worth to note that the same kind of approximation can be found
in solid state physics in the so called Anderson ``tight binding''
model~\cite{Anderson}, where the electronic wave function is assumed
to be localized around the nuclei.

In presence of reaction eq.~(\ref{eq:evt}) changes into
\begin{equation}
\theta_n(t+\Delta t)=G_{\Delta t} \left (
   \sum_j P^{(\Delta t)}_{j\to n} \theta_j(t) \right )\,,
\label{eq:discreterule}
\end{equation}
where $G_{\Delta t}(\theta)$ is an assigned reaction map.
For a discussion on how to obtain the previous rule from the
basic equation~(\ref{eq:ard}) see~\cite{acvv01,Mancinelli}.

The shape of the reaction map $G_{\Delta t}(\theta)$
depends on the underlying chemical model.
For an autocatalytic reaction (the FKPP class),
characterized by an unstable fixed point in 
$\theta = 0$ and a stable one in $\theta=1$, one has:
$
   G_{\Delta t}(\theta) = \theta + \theta(1-\theta) \Delta t /\tau.
$
For ignition-type class, instead, the reactive map reads:
\begin{equation}
   G_{\Delta t}(\theta) = \left \{ \begin{array}{lll}
   \theta &  & 0 \leq \theta \leq \theta_c \\
   \theta + (\theta - \theta_c)(1-\theta) 
    {\displaystyle {\Delta t \over \tau}} & 
    & \theta_c < \theta \leq 1.
    \end{array}
   \right .
\label{eq:ignition}
\end{equation}

We expect from known results~\cite{xin2000}, valid for the
time-continuous PDE (\ref{eq:ard}) that, at a qualitative level, the
detailed shape of $G_{\Delta t}(\theta)$ is not very relevant, within
a given class of nonlinearities ({\it e.g.}~FKPP or ignition-like).
This expectation is confirmed by numerical simulations.\\

\section{Numerical results}
Let us now present the results of numerical computations for the
system~(\ref{eq:discreterule}).  For the sake of simplicity we
consider a spacing $\Delta x=1$ (the distance between two fixed point
of eq.~(\ref{eq:lagrangian})); the lattice size being $L_x \leq 4
\cdot 10^4$. We use a time step $\Delta t \leq 10^{-2}$ and an initial
condition localized around $n = 0$, $\theta_n(0) = \Theta_n$, where
\begin{equation}
   \Theta_n = \left \{ \begin{array}{l}
   1 \qquad {\mathrm {for}} \,\,\, |n| \leq {\ell \over 2} \\
   0 \qquad {\mathrm {for}} \,\,\, |n| > {\ell \over 2}. 
    \end{array}
   \right .
\label{eq:initcond}
\end{equation}
A useful observable to focus on is the spatial integral of the scalar
field $\theta_n$, which represents the total burnt area in the case of
ideal fronts; therefore we compute its analogue on the lattice,
expressed by the quantity
\begin{equation}
Q(t) = \sum_{n=-\infty}^{+\infty} \theta_n(t)\,.
\label{eq:obs}
\end{equation}
In absence of quenching we have an asymptotic linear growth
of $Q(t)$, that is
\begin{equation}
Q(t) \simeq 2 v_f t \qquad \mbox{for large } t\,,
\label{eq:vel}
\end{equation}
where $v_f$ is the front speed.
The coefficient 2 is here due to the fact that with our choice for
the initial condition two symmetric fronts develop.
In the case of autocatalytic reaction term
we obtain (for large $\tau$ and $W$)
the expected result valid for the continuous
FKPP limit $v_f = 2 \sqrt{W/\tau}$.

\section{Ignition reaction term}
\begin{figure}[htbp]
\includegraphics[draft=false,scale=0.68,clip=true]{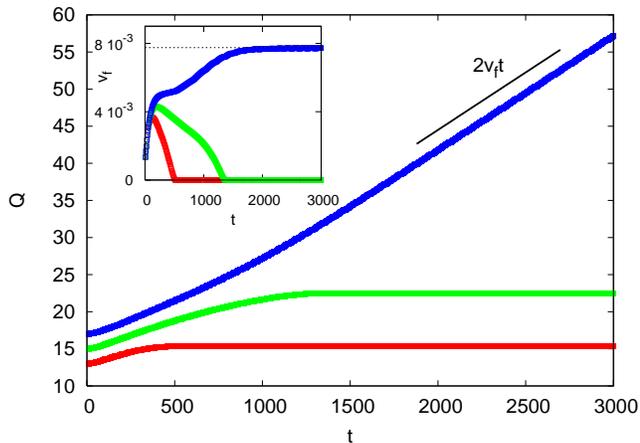}
\caption{(color online) $Q$ as a function of the time $t$ for 
$\theta_c=0.6$, $W=0.1$ and $\tau=50$. The initial condition widths are $\ell=14,16,18$ 
from bottom to top.  In the propagating case ($\ell=18$) 
$Q \simeq 2 v_f t$ for large times. Inset: front speed $v_f$ {\it vs.} $t$.}
\label{fig.1}
\end{figure}

Now we consider the ignition case with the reaction
term~(\ref{eq:ignition}) and investigate the possibility 
of quenching of the reactive dynamics.
This could occur for large values of the threshold density 
$\theta_c$ and/or for narrow initial conditions, and also
depends on the reaction time, $\tau$, and on the combined effects
of molecular diffusivity and advective flow, $W$.
As a first example, we show in Figure~\ref{fig.1}
the system dynamics at varying the initial
width $\ell$. The quenching appearance can be detected 
following the behaviour in time of $Q$ and $v_f$.
If the initial condition is narrow enough, 
after a transient the growth of $Q$ is arrested and correspondingly the front 
speed goes to zero. For larger values of $\ell$ propagation takes place with 
the asymptotic time behaviour $Q \simeq 2 v_f t$.
In such a way it is possible to determine a critical length $\ell_c$
separating the two regimes:
\begin{center}
\begin{tabular}{rcll}
$\ell < \ell_c$ & $\Rightarrow$ & $\theta(x,t \rightarrow \infty) \rightarrow 0$ 
& (quenching) \\
$\ell > \ell_c$ & $\Rightarrow$ & $\theta(x,t \rightarrow \infty) \rightarrow 1$ & 
(propagation).
\end{tabular}
\end{center}

\begin{figure}[htbp]
\includegraphics[draft=false,scale=0.68,clip=true]{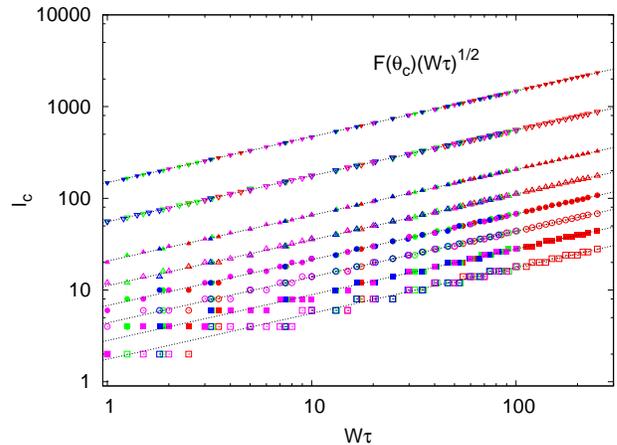}
\caption{(color online) Critical width of the initial condition as a
function of the product $W\tau$ for data sets coming from both
experiments A ($\tau=10, \tau=50$) and B ($W=0.1, W=0.25$). The
different colours correspond to the four different experimental
settings and symbol types to different values of $\theta_c$; from
bottom to top $\theta_c=0.3,0.4,...,0.9,0.95$.}
\label{fig.2}
\end{figure}

The critical value of the initial width will depend on the relevant
physical parameters of the problem: $W$, $\tau$ and $\theta_c$. In
order to investigate this point we perform two types of numerical
experiments. In the first one (experiment A) we keep the reaction time
$\tau$ fixed and vary the escape rate $W$ for a given set of values of
$\theta_c$. In the second one (experiment B), the situation is
reversed, namely, for the same values of $\theta_c$, we study
how $\ell_c$ varies with $\tau$ when $W$ is kept constant.
Irrespective of the specific value of the threshold concentration, in
both cases A and B we find a square-root relation between $\ell_c$ and
the product $W \tau$: \BE \ell_c=F(\theta_c)\sqrt{W\tau}
\label{eq:ellc}
\EE 
where $F(\theta_c)$ is a constant factor containing the
dependence on $\theta_c$ (see Figure~\ref{fig.2}). 

\begin{figure}[htbp]
\includegraphics[draft=false,scale=0.68,clip=true]{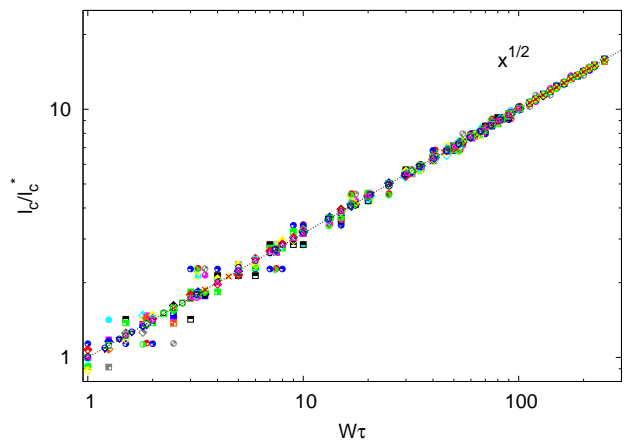}
\caption{(color online) Collapse of the data reported in Fig.~\ref{fig.2} showing 
the universality of the square-root law. In the vertical axis we plot
$\ell_c/\ell_c^*$ where $\ell_c^*$ is computed for $W\tau=1$.}
\label{fig.3}
\end{figure}

Relation (\ref{eq:ellc}) can be derived by a dimensional argument. 
In the continuum limit of the lattice model,{\it i.e.}, $L_x \gg \Delta x$,
the system can be regarded as a pure reaction-diffusion system with
diffusivity equal to $D=W\Delta x^2=W$, 
since we use $\Delta x = 1$. 
Then, the only possibility to build a length-scale
with the quantities $W \equiv D$, $\tau$ and $\theta_c$ is
$\sqrt{W\tau}F(\theta_c)$, where $F$ is a nondimensional function of
the threshold concentration.  If the initial width of the burnt area
is smaller than this, then the ``equivalent diffusion'',
{\it i.e.}, the combined effects of diffusion and velocity field,
will be efficient enough to
spread the majority of the inert material below the concentration threshold  
on a reactive time-scale and, consequently, to quench the
reaction.  The above results are summarized in
Figure~\ref{fig.2}, where $\ell_c$ is plotted against
$W\tau$, and Figure~\ref{fig.3} where all data are collapsed
onto a single curve showing the universality of the square-root
dependence. 

\begin{figure}[htbp]
\includegraphics[draft=false,scale=0.68,clip=true]{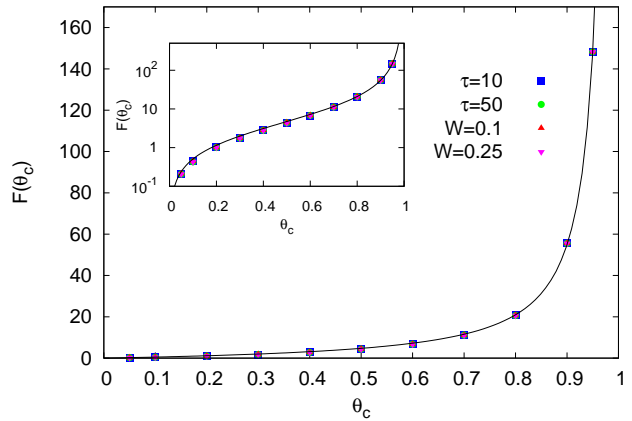}
\caption{(color online) Plot of the function $F(\theta_c)$ for
  experiments A ($\tau=10$, $\tau=50$) and B ($W=0.1$, $W=0.25$).  
  The solid line is a fit with a function corresponding to the second
  order expansion around the singularity $z=1$: $F(z)=a_{0}{z \over
  {1-z}}\left( 1+ {a_{-2} \over a_{0}} {1 \over {1-z}}\right)$; $a_{0}
  \simeq 4.39$, $a_{-2} \simeq 0.17$; for the piecewise-linear
  reaction map $\tilde{G}_{\Delta t}$ (see 
  eq.~(\ref{eq:ignition_triang}) in the text) $a_{0} \simeq 4.88$,
  $a_{-2} \simeq 0.20$. Inset: semilogarithmic plot of the same
  function.}
\label{fig.4}
\end{figure}

A natural question arises, concerning the shape of the function
$F(\theta_c)$ appearing in eq.~(\ref{eq:ellc}).  Its values, measured
in experiments of type A and B, are reported in
Figure~\ref{fig.4}. The perfect superposition of data corresponding to
different experimental settings reflects the robustness of the
dimensional estimate~(\ref{eq:ellc}), and the fact that the dependence
on $\theta_c$ can be found only in the prefactor, $F(\theta_c)$.\\ 
In order to clarify the dependence on $\theta_c$ we consider an {\it
ansatz} based on the following very general physical hypothesis:
\begin{description}
\item[(i)] $F(\theta_c)$ is a non-negative function, monotonically
  increasing with $\theta_c \in [0,1]$
\item[(ii)] $F(\theta_c) \rightarrow 0$ when $\theta_c \rightarrow 0$
\item[(iii)] $F(\theta_c) \rightarrow \infty$ when $\theta_c \rightarrow 1$
\item[(iv)] $F(\theta_c)$ is analytic for $\theta_c\neq 1$.
\end{description}

Some comments are in order. Hypothesis (ii) and (iii) correspond to
the physical expectation that when the threshold is very small the
reaction proceeds and when it is very large it quenches,
respectively. Moreover, when $\theta_c \rightarrow 0$ the system
clearly cannot exhibit quenching, since in that limit the reaction
term (\ref{eq:ignition}) reduces to the discrete-time version of the
autocatalytic FKPP term $G_{\Delta t}(\theta)=\theta+
\theta(1-\theta)\Delta t/\tau$, which is known to always give rise to front
propagation~\cite{Rocquejoffre1,Rocquejoffre2}.
Hypothesis (iv) states that the only singular point we expect is
$\theta_c=1$.

According to the above hypothesis, we can Laurent-expand the function
$F(z)$ around the point $z=1$: 
\begin{equation} F(z)=a_0+{a_{-1} \over
{1-z}}+{a_{-2} \over {(1-z)^2}}+... = \sum_{k=0}^\infty {a_{-k} \over
{(1-z)^k}}
\label{eq:laurent}
\end{equation}
The expansion will be
truncated at a certain order $\alpha$ if all the coefficients $a_{-k}$
with $k>\alpha$ are zero, that is if the singularity is a pole of
order $\alpha$. The numerics 
suggest that indeed the point $z=1$ is a pole of order $\alpha=2$. In
other words:

\BE
\lim_{z \rightarrow 1} (1-z)^\alpha F(z)=0 \quad \mbox{for} 
         \quad \alpha \geq 3
\label{eq:pole}
\EE
and therefore we conjecture the {\it ansatz}
\BE
F(z)=a_0+{a_{-1} \over {1-z}}+{a_{-2} \over {(1-z)^2}}.
\label{eq:ansatz}
\EE
Though this is formally a 3-parameter family of functions, one of the
parameters can be eliminated imposing the physical constraint
$F(z=0)=0$ (hypothesis (ii)). In the end, by doing so, we get the
following expression for $F(\theta_c)$:
\BE
F(z)=a_{0}{z \over {1-z}}\left( 1+ {a_{-2} \over a_{0}} {1 \over {1-z}}\right).
\label{eq:Fthetac}
\EE
In this form, the role of the extremal points $\theta_c=0,1$ is
evident: if $\theta_c \rightarrow 0$ then $F$ vanishes and so does
$\ell_c$, that is, propagation always prevails. On the contrary, when
$\theta_c \rightarrow 1$ the divergence of $F$ implies that of the
critical width of the initial condition, corresponding to quenching of
the reaction independently of the fixed values of $W$ and $\tau$.  
Therefore, in a practical situation, an improved estimate of the
scaling relation $\ell_c \sim \sqrt{W\tau}$ can be obtained by using
the heuristic expression~(\ref{eq:Fthetac}).
In Figure~\ref{fig.4} we report a comparison
between a fit with the function in eq.~(\ref{eq:Fthetac}) and the
numerical results; the agreement is rather good, confirming our
conjecture.

In order to check the robustness of the above result 
we considered another ignition reaction map in place of 
eq.~(\ref{eq:ignition}) 
\begin{equation}
   \tilde{G}_{\Delta t}(\theta) = \left \{ \begin{array}{lr}
   \theta  & 0 \leq \theta \leq \theta_c \vspace{0.2truecm}\\
   \theta + {\displaystyle {{1-\theta_c} \over {2 \tau}}}
             \Delta t (\theta - \theta_c)
    & \theta_c < \theta \leq \theta_* \vspace{0.2truecm}\\
   \theta + {\displaystyle {{1-\theta_c} \over {2 \tau}}}
             \Delta t  (1-\theta) 
    & \theta_* < \theta \leq 1
    \end{array}
   \right .
\label{eq:ignition_triang}
\end{equation}
where $\theta_*=(1+\theta_c)/2$. Numerical simulations indeed
demonstrate (results not shown) that the scaling behaviour
of $\ell_c$ and the shape of the function $F(\theta_c)$ do not 
significantly change.

It is natural to wonder about the existence (or not) of a link between
the critical length $\ell_c$ and the characteristic front thickness
$\xi \propto \sqrt{W \tau}$.  We remind that $\xi$ can be defined from
the asymptotic shape of the propagating front. In order to guarantee
front propagation one can assume as initial condition $\theta=1$ for
$x<0$ ($x>0$ for the symmetric case), which implies an infinite
reservoir of burnt material.  
A first question is whether the front, in the case of a compressible
velocity field and with ignition reaction term, has a different shape
from that of the paradigmatic FKPP model. It is known from theoretical
results (see, {\it e.g.}, \cite{Aronson}) that the standard FKPP front
shape is exponential, {\it i.e.}, for $x \gtrsim v_0 t$ one has
$\theta(x,t)\sim \exp[-(x-v_0t)/\xi_0]$, where
  $v_0=2\sqrt{D/\tau}$ and $\xi_0=\sqrt{D\tau}$ 
are the FKPP front speed and length, respectively.

\begin{figure}[htbp]
\includegraphics[draft=false,scale=0.68,clip=true]{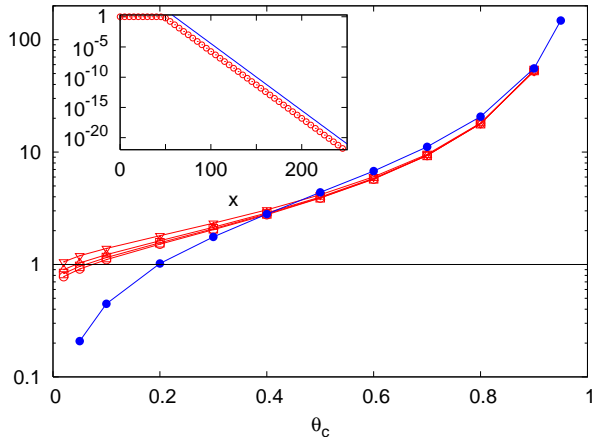}
\protect\caption{(color online) Plot of the rescaled front thickness and the function
$F(\theta_c)$. 
The symbols $\square$, $\circ$, $\bigtriangledown$ and
$\bigtriangleup$ indicate the rescaled front thickness in the cases 
$W=0.1$ and $\tau=10$, $W=0.25$ and $\tau=10$, $W=0.25$ and $\tau=2$, 
$W=0.1$ and $\tau=2$, respectively. The symbol $\bullet$ 
indicates the function $F(\theta_c)$. The straight line is the
$\xi_0$ value in the case of a pure FKPP process.
In the inset it is shown the right side of the front shape.}
\label{fig.5}
\end{figure}

In the inset of Figure~(\ref{fig.5}) the shape of the right
propagating front is shown. Its exponential shape is well evident.
This result allows us to use the following expression for the front shape:
\begin{equation}
  \theta(x,t) \propto \exp\left (-{x-v_f t\over \xi} \right )\,,
  \label{eq:xidef}
\end{equation}
from which the front length $\xi$ can be computed.\\
To investigate the link between $\xi$ and
$\ell_c$ we measured the front length at varying $\theta_c$.
In Figure~\ref{fig.5} it is possible to observe that for
$\theta_c \gtrsim \theta^* \approx 0.3$ one has $\ell_c \approx \xi$.
On the contrary, for $\theta_c \lesssim \theta^*$, $\ell_c$ is smaller
than $\xi$. In particular for very small values of $\theta_c$ one has
$\ell_c \to 0$ while $\xi \to \xi_0 = \sqrt{W \tau} \neq 0$.
This is indicative of the fact that the quenching phenomenon is not simply
related to the (usual) features of the front.

\section{Conclusions}
Let us now conclude with some general considerations and a comparison
of our results with others obtained for incompressible bidimensional
velocity fields.\\
As first, we note that for $\Delta x \to 0$ the
rule~(\ref{eq:discreterule}) is, with a suitable rescaling of the
parameters, nothing but the finite difference discretization algorithm
to solve eq.~(\ref{eq:ard}) with $\bu = 0$. Therefore, our numerical
results are also related to the quenching problem of the pure
reaction-diffusion system with ignition-like nonlinearities.
For the latter case there exists a theoretical prediction of the
system behaviour~\cite{Kanel,Zlatos} that is in good agreement with
our results.

Moreover, in refs.~\cite{Constantin1,Constantin2} Constantin and
co-workers performed detailed numerical simulations of the quenching
problem in the case of slow reaction in two-dimensional incompressible
velocity fields, in particular for
\begin{itemize}
\item[a)] shear flow of typical intensity $U$,
\item[b)] cellular flow of typical intensity $U$,
\end{itemize}
obtaining $\ell_c \sim U$ in case a) and $\ell_c \sim U^{1\over4}$
in case b). Such a conclusion can be easily {related to} our
results. In fact, in the slow reaction limit the long time and large
scale behaviour of~(\ref{eq:ard}) can be written as
\begin{equation}
\partial_t \theta = D^{\mathrm {eff}}\nabla \theta+{1\over\tau}f(\theta)
\label{eq:ardrenorm}
\end{equation}
where $D^{\mathrm {eff}}$ depends (often in a non-trivial way) on the
velocity field $\bu$ (see, {\it e.g.}, \cite{acvv01}). Therefore, we
can use the previous result (on the connection 
between~(\ref{eq:discreterule}) and the pure reaction-diffusion
problem without velocity field) and conclude that $\ell_c \sim
\sqrt{D^{\mathrm {eff}}\tau}$. Using the well known result (see, {\it
e.g.}, \cite{acvv01}) that $D^{\mathrm {eff}} \sim U^2$ for the shear
flow (case a)) and $D^{\mathrm {eff}} \sim U^{1\over 2}$ for the
cellular flow (case b)) one obtains the result of Constantin {\it et
al.} \cite{Constantin1,Constantin2}.

In conclusion, we studied the quenching phenomenon of ignition-type
reaction dynamics in a steady compressible flow. We developed a
simplified lattice model based on a physically controllable
localization approximation for the concentration field, which allows
an efficient numerical implementation. The dependence of the critical initial
condition width $\ell_c$ on the relevant parameters $W,\tau,\theta_c$
was established by means of numerical experiments and dimensional
reasoning. Finally we compared our results with those obtained 
theoretically and numerically in different flow configurations.

\acknowledgments 
SB acknowledges financial support from CNRS and partial support from
TEKES during the early stage of this work.


\end{document}